\documentclass[twocolumn,aps,prc,showpacs,superscriptaddress,preprintnumbers,amssymb,floatfix,nofootinbib]{revtex4}

\usepackage{xspace}
\usepackage[colorlinks=true,linkcolor=blue, citecolor=blue]{hyperref}

\usepackage{bm}
\usepackage{mathrsfs}
\usepackage{float}
\usepackage{graphicx}
\usepackage{epstopdf}
\usepackage{bbold}
\usepackage{comment}
\usepackage{color}
\usepackage{soul}
\usepackage[abs]{overpic}
\usepackage{amsmath}
\usepackage{lipsum}
\usepackage[abs]{overpic}
\usepackage{nicefrac}
\usepackage{subfigure}
\usepackage{gensymb}

\usepackage[normalem]{ulem}  

\renewcommand\sout{\bgroup \color{red} \ULdepth=-.5ex \ULset}

\begin{document}

\title{Hadronization of $\Lambda_c^+$ baryons from recombination model in Pb+Pb collisions at the Large Hadron Collider}
    \author{Jing-Zong Zhang}
    \affiliation{College of Physics, Sichuan University, Chengdu 610064, China}
\affiliation{Key Laboratory of Nuclear Physics and Ion-beam Application (MOE), Institute of Modern Physics, Fudan University, Shanghai 200433, China}
	\affiliation{Shanghai Research Center for Theoretical Nuclear Physics, NSFC and Fudan University, Shanghai $200438$, China}
\author{Hua Zheng}\email{zhengh@snnu.edu.cn}
\affiliation{School of Physics and Information Technology, Shaanxi Normal University, Xi'an 710119, China}
\author{Lilin Zhu}\email{zhulilin@scu.edu.cn}
\affiliation{College of Physics, Sichuan University, Chengdu 610064, China}

\begin{abstract}
The production of $\Lambda_c^+$ baryons in Pb+Pb collisions at $\sqrt{s_{NN}} = 5.02$ TeV is investigated within the quark recombination framework, including the energy loss of light and charm quarks inside the hot and dense medium. The model simultaneously describes the transverse momentum ($p_T$) spectra of  $\Lambda_c^+$ baryons, $\Lambda_c^+/D^0$ yield ratio, which is attributed to the dominance of quark recombination mechanism in the Quark-Gluon Plasma (QGP), and the second harmonic coefficient of $\Lambda_c^+$ baryons with emphasis on the effects of minijets on the azimuthal anisotropy. Furthermore, we extend the theoretical calculation to Pb+Pb collisions at $\sqrt{s_{NN}} =2.76$ TeV and make predictions for $\Lambda_c^+$ baryons and $\Lambda_c^+/D^0$ yield ratio. The simultaneous description of the yield, baryon-to-meson ratio, and azimuthal anisotropy further validates that the recombination model is an effective hadronization mechanism in heavy-ion collisions. 

\end{abstract}

\maketitle

\section{Introduction}
Ultra-relativistic heavy-ion collisions provide a unique laboratory for probing quantum chromodynamics (QCD) under extreme conditions of temperature and energy density \cite{Busza:2018rrf, Bazavov:2011nk}. The hadronization of heavy quarks constitutes one of the most compelling probes of the Quark-Gluon Plasma (QGP) created in ultra-relativistic heavy-ion collisions. Due to their large masses, heavy quarks are predominantly produced in the initial hard scatterings and subsequently propagate through the strongly interacting medium. During the past decade, experimental advances at the Relativistic Heavy Ion Collider (RHIC) and the Large Hadron Collider (LHC) have enabled the precise measurement of not only open-charm mesons but also charmed baryons \cite{STAR:2018zdy, STAR:2017kkh, ALICE:2021rxa, ALICE:2021kfc, ALICE:2023gco, ALICE:2018hbc, STAR:2019ank,ALICE:2021bib, DiCostanzo:2025zyb}. 
A striking observation is the significant enhancement of the $\Lambda_c^+/D^0$ yield ratio at intermediate transverse momentum in Pb+Pb collisions at $\sqrt{s_{NN}} = 5.02$ TeV \cite{ALICE:2021bib}, which contradicts the expectation from vacuum fragmentation functions and resembles the ‘‘baryon anomaly’’ observed for light flavors such as $p/\pi$ and $\Lambda/K_s^0$ \cite{PHENIX:2003iij, ALICE:2019hno, ALICE:2013cdo}. This enhancement can be used to probe the in-medium hadronization dynamics of heavy quarks and put stronger constraint on the heavy quark hadronization mechanism in heavy-ion collisions.

Several theoretical approaches have been developed to describe the production of charmed hadrons in heavy-ion collisions, such as the statistical model \cite{Andronic:2002pj, Andronic:2021erx}, transport model \cite{Plumari:2017ntm, He:2019vgs}, and quark recombination or coalescence (ReCo) model \cite{Zhao:2023nrz, Greco:2003vf, Cho:2019lxb, Zhang:2025ehd}. Especially, the ReCo model offers a natural explanation for baryon-to-meson enhancement and has also successfully described the production of light hadrons \cite{hy, vg,rf}. Furthermore, its extension to charm quarks has attracted significant attention.  More than a decade ago,  ReCo model was already employed to study the heavy flavor hadron ratios $\Lambda_c^+/D^0$ and $\Lambda_b/\bar{B}^0$ produced in central Au+Au collisions at $\sqrt{s_{NN}} = 200$ GeV \cite{Oh:2009zj}. Although several theoretical approaches exist, a unified framework that simultaneously describes the transverse momentum spectra, $\Lambda_c^+/D^0$ yield ratio and the azimuthal anisotropy across different centralities within a single set of physical parameters remains a challenge. In this work, we present a systematic study of the production of $\Lambda_c^+$ baryons in Pb+Pb collisions at $\sqrt{s_{NN}} = 5.02$ TeV within the quark recombination model. Our approach explicitly incorporates the recombination of thermal partons with shower partons generated by energetic partons losing energy in the hot medium. We demonstrate that this mechanism not only reproduces the transverse momentum distributions of $\Lambda_c^+$ baryons and $\Lambda_c^+/D^0$ yield ratio but also provides a natural explanation for the observed second harmonic of azimuthal anisotropy $v_2$ of $\Lambda_c^+$. Our results emphasize the essential role of minijets in bridging the soft and hard sectors of the collision system and support the dominance of recombination in the hadronization of charm quarks produced in heavy-ion collisions.

This paper is organized as follows. In Sec. \ref{model}, we introduce the recombination model which incorporates both thermal and shower parton distributions, and present the formalism for $\Lambda_c^+$ baryons via parton recombination. In Sec. \ref{results}, we establish the theoretical calculations of transverse momentum spectra of $\Lambda_c^+$ baryons at central and mid-central collisions, $\Lambda_c^+/D^0$ yield ratio and the second harmonic coefficient of $\Lambda_c^+$ baryons in Pb+Pb collisions at $\sqrt{s_{NN}} = 5.02$ TeV. These results are compared with available experimental data. The predictions for the transverse momentum spectra of $\Lambda_c^+$ baryons and $\Lambda_c^+/D^0$ yield ratio in Pb+Pb collisions at $\sqrt{s_{NN}} = 2.76$ TeV are also provided, without introducing any additional free parameters. A summary and outlook are given in Sec. \ref{summary}.
 
\section{Quark recombination for $\Lambda_c^+$ baryons}
\label{model}
In the quark recombination model, hadron production is described as the coalescence of constituent quarks that are close in phase space. For baryons composed of quarks ($q_1, q_2, q_3$) with momenta ($p_1, p_2, p_3$), the invariant distribution at midrapidity can be expressed as \cite{Zhu:2014csa, Zhu:2021fbs}
\begin{equation}
p^0\frac{dN^B}{dp_T} =\int \prod_{3}^{i=1}\frac{dp_i}{p_i} F_{q_1q_2q_3}(p_1,p_2,p_3)R^B_{q_1q_2q_3 }(p_1,p_2,p_3,p_T),
\label{2.1}
\end{equation}
$R_{q_1q_2q_3 }^B$ is the recombination function (RF) for baryons. For $\Lambda_c^+$, we have the following parametrization \cite{Anjos:2001jr, Herrera:1997qh}
\begin{equation}
R_{\Lambda_c^+}(p_1,p_2,p_3,p_T) =  g_{st}^{\Lambda_c^+}\frac{p_1p_2p_3}{p_T^3}
\delta\left(\frac{p_1}{p_T}+\frac{p_2}{p_T}+\frac{p_3}{p_T}-1\right),
\label{2.2}
\end{equation}
where $g_{st}^{\Lambda_c^+}$ is the statistical factor and the delta function enforces momentum conservation. 
$F_{q_1q_2q_3}$ is the parton distribution before hadronization. In the recombination model, the partons are divided into two types: thermal (T) and shower (S). Taking into account the recombination of different types of partons, we thus have
\begin{eqnarray}
F_{q_1q_2q_3} &=& \mathcal{T}\mathcal{T}\mathcal{T}+\mathcal{T}\mathcal{T}\mathcal{S}+\mathcal{T}\mathcal{S}\mathcal{S}+\mathcal{S}\mathcal{S}\mathcal{S},
\label{2.3}
\end{eqnarray}
where $\mathcal{T}$ and $\mathcal{S}$ are used to denote the invariant distributions for thermal and shower partons, respectively. For a visualization of the various components, the schematic diagrams are presented in Ref. \cite{Zhu:2014csa}.  The shower partons in $\mathcal{T}\mathcal{S}\mathcal{S}$ can come from one or two jets. The case for $\mathcal{S}\mathcal{S}\mathcal{S}$ is more complicated. In this work, only the contributions from $\mathcal{S}\mathcal{S}\mathcal{S}^{1j}$ and $\mathcal{S}\mathcal{S}\mathcal{S}^{2j}$ are retained, since the component $\mathcal{S}\mathcal{S}\mathcal{S}^{3j}$, which corresponds to three shower partons originating from three distinct jets, is strongly suppressed.

The soft partons generated by multiple partonic scattering and radiation in the medium interact with the bulk partons, and cannot be distinguished from the latter by the time the density of all soft partons is low enough for hadronization. They are all referred to here as thermal partons in the final stage of the quark matter as they move out of the deconfinement phase. The thermal parton distribution is assumed to follow a simple exponential form \cite{Zhu:2014csa, Zhu:2021fbs}
\begin{equation}
\mathcal{T}_j(p_i) = p_i\frac{dN_j}{dp_i} = C_j p_i e^{-p_i/T_j},
\label{2.4}
\end{equation}
where $C_j$ and $T_j$ are the normalization constant and inverse slope parameter, respectively, which have already been determined in our previous study on the production of charmed mesons in Pb+Pb collisions at $\sqrt{s_{NN}} = 5.02$ TeV \cite{Zhang:2025ehd}. Their values are listed in TABLE \ref{tab1} for the convenience of readers.

\begin{table}
\tabcolsep0.3in
\renewcommand{\arraystretch}{1.3}
\begin{tabular}{ccc}
\hline
\hline
Centrality &0-10\% &30-50\% \\ 
 \hline
$C_q$ [(GeV/c)$^{-1}$] & 23.8 & 15 \\
$C_c$ [(GeV/c)$^{-1}$] &1.1 &0.39 \\
$T_q$ (GeV) &0.42 &0.42\\
$T_c$ (GeV) &0.79 &0.79\\
 \hline
 \hline
 \end{tabular}
 \caption{Parameters $C_q$, $C_c$, $T_q$ and $T_c$  in Eq. (\ref{2.4})  for Pb+Pb collisions at $\sqrt{s_{NN}}=5.02$ TeV.} 
 \label{tab1}
 \end{table}
 
\begin{table*}
\tabcolsep0.3in
\begin{tabular}{ccccc}
\hline\hline
 $\sqrt{s_{NN}}$ (TeV)  & &$A$ ($10^4$/GeV$^2$) & $B$ (GeV) & $n$ \\
 \hline
 &$g$  &  11.2 & 0.80 &5.68 \\
 &$u$  &  2.02 &0.59 &5.31 \\
5.02& $d$  & 2.28  &0.58  &5.29 \\
 &$\bar u$  &  0.42 &0.75  &5.52 \\
 &$\bar d$   &0.40 &0.76  &5.53 \\
 &$s, \bar s$ &0.154 &0.93 &5.63 \\
 \hline \hline
 \end{tabular}
 \caption{Parameters $A$, $B$ and $n$ for $f_i(k, 0.05)$ in Eq.\ (\ref{2.10}) for central Pb+Pb collisions at $\sqrt{s}=5.02$ TeV \cite{Zhu:2021fbs}.}
 \label{tab2}
 \end{table*}
 
As discussed in Refs. \cite{Zhu:2021fbs, Zhang:2025ehd}, the shower parton distribution at centrality $c$ is written as
\begin{equation}
\mathcal{S}^j(p, c) = \int \frac{dq}{q}\sum_i \hat{F}_i(q, c)S_i^j(p, q),
\label{2.5}
\end{equation}
where $S_i^j(p, q)$ is the shower parton distribution (SPD) of a parton of type $j$ fragmenting from a jet of type $i$ with momentum fraction $p/q$, determined through the fragmentation function (FF) as described in Refs. \cite{hy4, Peng:2010zza}. $\hat{F}_i(q, c)$ is given by
\begin{equation}
\hat{F}_i(q,c) = \frac{1}{2\pi}\int d\phi \int d\xi P_i(\xi,\phi,c)\int dk k f_i(k,c) G(k,q,\xi),
\label{2.6}
\end{equation}
where $G(k,q,\xi)$ represents the momentum degradation due to parton energy loss parametrized as \cite{Zhu:2014csa}
\begin{equation}
G(k,q,\xi) = q\delta(q-ke^{-\xi}),
\label{2.7}
\end{equation}
relating the initial parton momentum $k$ to the final momentum $q$ at the medium surface by an exponential decay via the dynamical path length $\xi$. $P_i(\xi,\phi,c)$ denotes the probability for parton $i$ to traverse the path length $\xi$ at an azimuthal angle $\phi$ in collisions with centrality $c$. The probability is initiated from position $(x_0, y_0)$ and weighted by the nuclear overlap function integrated over all $(x_0, y_0)$. The intrinsic connection between the collisions geometry and the parton dynamics is imbedded in $P_i(\xi,\phi,c)$, where the dynamical path length $\xi$ is proportional to the geometrical path length $l$. The detailed explanation of $l$ have been presented in our earlier work \cite{Zhu:2014csa} and will not be repeated here. Hence, $P_i(\xi,\phi,c)$ can be calculated by \cite{Chiu:2008ht}
\begin{equation}
P_i(\xi,\phi, c) = \int dx_0dy_0Q(x_0,y_0,c)\delta(\xi-\gamma_il),
\label{2.8}
\end{equation}
where $Q(x_0,y_0,c)$ is the probability for a hard (or semihard) parton to be produced at $(x_0, y_0)$. The factor $\gamma_i$ accounts for the jet-quenching effects in the medium, which enhance the parton energy degradation due to soft parton creation. For Pb+Pb collisions, $\gamma_g$ for gluons is parameterized as \cite{Zhu:2014csa}
\begin{equation}
\gamma_g(q) =  \frac{\gamma_0}{1+(q/q_0)^2},
\label{2.9}
\end{equation}
with $q_0=7$ GeV/c and $\gamma_0=4.5$ for gluons and $1.0$ for charm quarks, as determined in our previous works \cite{Zhu:2021fbs, Zhang:2025ehd} for the production of light hadrons and charmed mesons in Pb+Pb collisions at $\sqrt{s_{NN}}=5.02$ TeV. As discussed before, light quarks ($i=u, d, s$) are assumed to undergo approximately half the energy loss experienced by gluons \cite{Zhu:2014csa}.  Charm quarks are produced in the initial hard-scattering process and their annihilation rate is small, so  the effect of jet quenching in the medium for charm quarks is smaller than that for light quarks. 
 
For light quarks, the centrality dependence of the minijet distribution $f_i(k,c)$ is assumed to be \cite{Zhu:2021fbs}
\begin{equation}
f_i(k,c) =\frac{T_{AA}(c)}{T_{AA}(0.05)} \times K\frac{A}{(1+k/B)^n},
\label{2.10}
\end{equation}
where $K=2.5$ and $c=0.05$ represents the centrality of 0-10\%. $T_{AA}$ is the nuclear thickness function, which accounts for shadowing effects and spatial fluctuations \cite{ALICE:2013hur}. The parameters $A$, $B$ and $n$ for Pb+Pb collisions at 5.02 TeV have been obtained by logarithmic interpolations of the parameters $\ln A$, $B$ and $n$ between Au+Au collisions at 200 GeV and Pb+Pb collisions at 5.5 TeV \cite{Zhu:2014csa, Zhu:2021fbs}, which are shown in TABLE \ref{tab2}. As done in our previous work \cite{Zhang:2025ehd}, the initial momentum distribution of charm quarks for $p + p$ collisions at $\sqrt{s_{NN}} = 5.02$ TeV can be extracted from PYTHIA8 \cite{Bierlich:2022pfr} simulations using the Monash tune. Then for charm quarks produced in Pb+Pb collisions, we employ the nuclear thickness function $T_{PbPb}$ for each centrality \cite{dEnterria:2003xac}:
\begin{equation}
\frac{d^2N^{PbPb}(c)}{dp_T dy} =T_{PbPb}(c)\frac{d^2\sigma^{pp}}{dp_T dy}.
\label{2.12}
\end{equation}

From the essential points discussed above, the thermal-thermal-thermal (TTT), thermal-thermal-shower (TTS), thermal-shower-shower (TSS), and shower-shower-shower (SSS) recombination for $\Lambda_c^+$ baryons can be written as 

\begin{widetext}
\begin{eqnarray}
\frac{dN^{TTT}_{\Lambda_c^+}}{p_Tdp_T}={g_{st}^{\Lambda_c^+} C_q^2 C_c\over m_T^{\Lambda_c^+} p_T^{2\alpha+\beta}} \int_0^{p_T}dp_1 \int_0^{p_T-p_1} dp_2 p_1 p_2 e^{-(p_1+p_2)/T_q} (p_T -p_1-p_2) e^{-(p_T -p_1-p_2)/T_c}, 
\label{2.12}
\end{eqnarray}
\begin{eqnarray}
\frac{dN^{TTS}_{\Lambda_c^+}}{p_Tdp_T}&=&{g_{st}^{\Lambda_c^+} \over m_T^{\Lambda_c^+} p_T^{2\alpha+\beta}} \int_0^{p_T}dp_1 \int_0^{p_T-p_1}dp_2
\left\{C_q^2 p_1p_2 e^{-(p_1+p_2)/T_q}  \mathcal{S}^c (p_T-p_1-p_2, c)\right.\nonumber\\
&&\left.+2C_qC_cp_1 e^{-p_1/T_q} p_2e^{-p_2/T_c} \mathcal{S}^u(p_T-p_1-p_2, c) \right\} , 
\label{2.13}
\end{eqnarray}
\begin{eqnarray}		
\frac{dN^{TSS^{1j}}_{\Lambda_c^+}}{p_Tdp_T}&=&{g_{st}^{\Lambda_c^+} \over m_T^{\Lambda_c^+} p_T^{2\alpha+\beta}} \int_0^{p_T}dp_1 \int_0^{p_T-p_1}dp_2
\left\{ C_cp_1e^{-p_1/T_c}\mathcal{S}^{ud}(p_2,p_T-p_1-p_2, c)\right.\nonumber\\
&&\left.+2C_qp_1 e^{-p_1/T_q}  \mathcal{S}^{dc}(p_2,p_T-p_1-p_2, c)\right\}, 
\label{2.14}
\end{eqnarray}
\begin{eqnarray}		
\frac{dN^{TSS^{2j}}_{\Lambda_c^+}}{p_Tdp_T}&=&{g_{st}^{\Lambda_c^+} \Gamma \over m_T^{\Lambda_c^+} p_T^{2\alpha+\beta}} \int_0^{p_T}dp_1 \int_0^{p_T-p_1}dp_2\left\{ C_cp_1e^{-p_1/T_c}\mathcal{S}^{u}(p_2, c)\mathcal{S}^{d}(p_T-p_1-p_2, c)\right. \nonumber\\
&&\left.+2C_qp_1 e^{-p_1/T_q}  \mathcal{S}^{d}(p_2, c)\mathcal{S}^{c}(p_T-p_1-p_2, c)\right\},
\label{2.15}
\end{eqnarray}
\begin{eqnarray}        
\frac{dN^{SSS^{1j}}_{\Lambda_c^+}}{p_Tdp_T}=\frac{1}{m_T^{\Lambda_c^+}}\int\frac{dq}{q^2}\hat{F}(q, c)D_c^{\Lambda_c^+}(p_T,q),
\label{2.16}
\end{eqnarray}
\begin{eqnarray}           
 \frac{dN^{SSS^{2j}}_{\Lambda_c^+}}{p_Tdp_T}&=&{g_{st}^{\Lambda_c^+} \Gamma \over m_T^{\Lambda_c^+} p_T^{2\alpha+\beta}} \int_0^{p_T}dp_1 \int_0^{p_T-p_1}dp_2 \left\{ \mathcal{S}^{c}(p_1, c)\mathcal{S}^{ud}(p_2,p_T-p_1-p_2, c)\right. \nonumber\\
&&\left.+  2{S}^{u}(p_1, c)\mathcal{S}^{dc}(p_2,p_T-p_1-p_2, c)\right\},
\label{2.17}
\end{eqnarray}
where
\begin{eqnarray}
\mathcal{S}^{qq}(p_2, p_3, c)=\int\frac{dq}{q}\sum\limits_i\hat F_i(q, c){\rm S}_i^q(p_2, q, c){\rm S}_i^q(p_3, q-p_2, c).
\label{2.18}
\end{eqnarray}
\end{widetext}

The shower-shower-shower recombination from one jet is equivalent to fragmentation process, so we can use fragmentation functions (FFs) $D_i^{\Lambda_c^+}$ in Eq. (\ref{2.16}). Following Ref. \cite{Delpasand:2020vlb}, when calculating the component $SSS^{1j}$, we only consider the contribution from charm quark $D_c^{\Lambda_c^+}$. The corresponding FFs for gluon and light quarks are set to zero at the initial scale \cite{Delpasand:2020vlb}. In Eqs. (\ref{2.15}) and (\ref{2.17}), $\Gamma$ is the probability of recombination between two parallel partons originating from different jets. Given that these partons are emitted from the medium at early times, the emitting system can be approximated as a thin, almond-shaped overlap region viewed edge-on in the transverse plane at midrapidity. From this perspective, the source effectively constitutes a one-dimensional system with a width of $\sim 10$ fm. Within this framework, the two parallel shower partons are treated as points on this line separated by a distance not greater than the diameter of a hadron ($\sim1$ fm). Consequently, we estimate $\Gamma=0.1$, which corresponds to the ratio of hadron diameter to nuclear diameter \cite{Zhu:2014csa, Zhu:2021fbs}. 

\section{Results and Discussion}
\label{results}
In this section, we show the results on the transverse momentum spectra of  $\Lambda_c^+$ baryons, $\Lambda_c^+/D^0$ yield ratio and the second coefficient of azimuthal anisotropy $v_2$ for $\Lambda_c^+$ baryons in Pb+Pb collisions at  $\sqrt{s_{NN}} = 5.02$ TeV. The inverse slopes $T_q$ and $T_c$ as well as the normalization factors $C_q$ and $C_c$ for Pb+Pb collisions at  $\sqrt{s_{NN}} = 5.02$ TeV determined in Ref. \cite{Zhang:2025ehd}, are listed in TABLE. \ref{tab1}. The values of parameters $\gamma_0$ and $q_0$ in Eq. (\ref{2.9}) are also shown in Sec. \ref{model}. Therefore, the only one unknown parameter is the statistical factor of $g_{st}^{\Lambda_c^+}$, which will be determined by the transverse momentum spectra of $\Lambda_c^+$ baryons. 

\subsection{Transverse momentum spectra and $\Lambda_c^+/D^0$ ratio}
\begin{figure}[pht]
\includegraphics[width=0.45\textwidth]{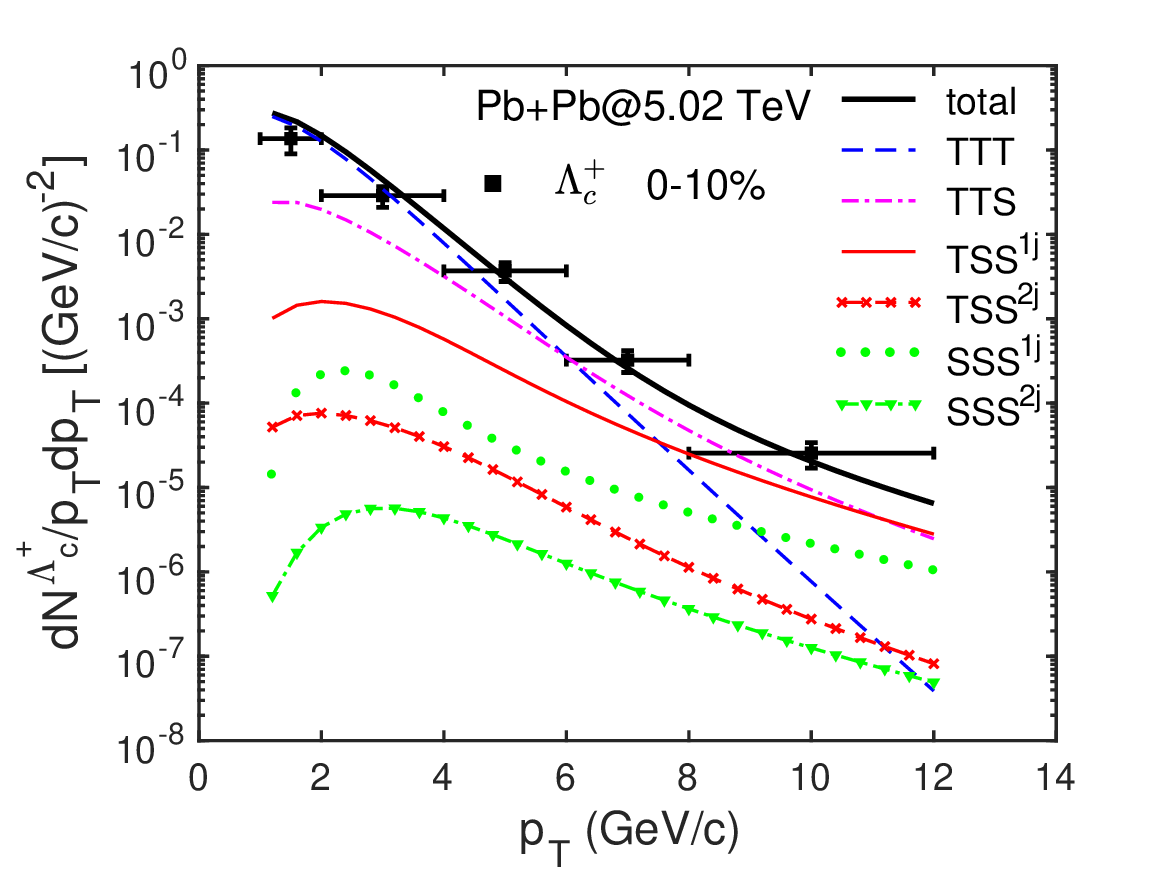}
 \caption{(Color online) Transverse momentum spectrum of $\Lambda_c^+$ baryons at centrality of 0-10\% in Pb+Pb collisions at $\sqrt{s_{NN}}=5.02$ TeV. The data are taken from Ref. \cite{ALICE:2021bib}.}
 \label{fig1}
\end{figure}

\begin{figure}[pht]
\includegraphics[width=0.45\textwidth]{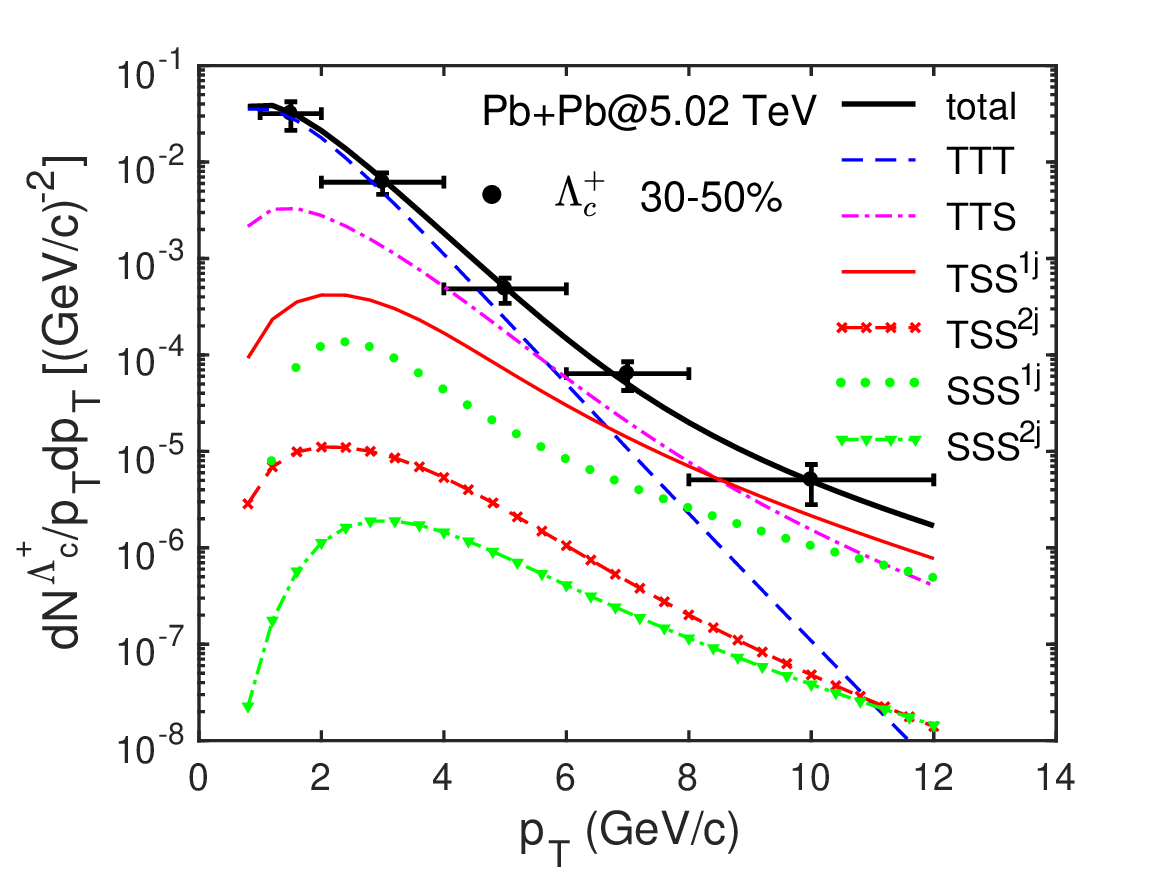}
 \caption{(Color online) Transverse momentum spectrum of $\Lambda_c^+$ baryons at centrality of 30-50\% in Pb+Pb collisions at $\sqrt{s_{NN}}=5.02$ TeV. The data are taken from Ref. \cite{ALICE:2021bib}.}
 \label{fig2}
\end{figure}

The transverse momentum spectrum of  $\Lambda_c^+$ baryons produced in Pb+Pb collisions at $\sqrt{s_{NN}} = 5.02$ TeV and the centrality of 0-10\% at midrapidity is shown by black line in Fig. \ref{fig1}. The recombination model reproduces very well the experimental data from the ALICE Collaboration shown by black solid squares \cite{Zhang:2025ehd}. The thermal and shower partons in various combinations are shown by different line types, which dominate at different $p_T$ regions. At low and intermediate $p_T$, the $\Lambda_c^+$ yield is dominated by the $\mathcal{TTT}$ component. The $\mathcal{TTS}$ and $\mathcal{TSS}$ components contain the jet-medium interaction at the hadronization stage, which is important at the region of $p_T>4$ GeV/c. $\mathcal{S}\mathcal{S}\mathcal{S}^{1j}$ recombination becomes important at high $p_T$, which is equivalent to fragmentation. At the same time, the contributions from two jets $\mathcal{T}\mathcal{S}\mathcal{S}^{2j}$ and $\mathcal{S}\mathcal{S}\mathcal{S}^{2j}$ are relatively small and can be neglected in the current analysis. The smooth transition from soft to hard regimes demonstrates that the unified recombination framework can consistently describe $\Lambda_c^+$ production across the entire $p_T$ range. A similar result is obtained for the transverse momentum distribution of  $\Lambda_c^+$ baryons at the centrality of 30-50\%. The agreement between the calculation results from the recombination model and the experimental data is quite good, as shown in Fig. \ref{fig2}. It should be emphasized that all parameters are fixed in our earlier work except the statistical factor $g_{st}^{\Lambda_c^+}$, whose value is equal to 0.24.

In our earlier work \cite{Zhang:2025ehd}, the production of $D^0$ mesons in Pb+Pb collisions at $\sqrt{s_{NN}} = 5.02$ TeV has been studied within the recombination model. Therefore, it is natural to consider the $\Lambda_c^+ / D^0$ yield ratio, which serves as a critical probe for testing our model. Figure 3 displays the $\Lambda_c^+ / D^0$ yield ratio as a function of transverse momentum in central (0-10\%) and mid-central (30-50\%) Pb+Pb collisions at $\sqrt{s_{NN}} = 5.02$ TeV. The experimental data from ALICE Collaboration are represented by black solid squares (0–10\%) and blue open circles (30–50\%) \cite{ALICE:2021bib}, while the results of quark recombination model are shown by solid black line (0–10\%) and short-dashed blue line (30–50\%), respectively. It is established that the recombination model can nicely describe the $\Lambda_c^+ / D^0$ yield ratio in mid-central collisions, which confirms that the QGP density at mid-central collisions optimally supports charmed baryon enhancement via quark recombination. The situation for central collisions is a little bit different. The theoretical calculation for central collisions provides a good description of the ratio at $p_T>4$ GeV/c, but overestimates at $p_T<4$ GeV/c. This is expected, as the $D^0$ yield did not include the contributions from resonance decays in Ref. \cite{Zhang:2025ehd}. Consequently, the calculated transverse momentum distribution for $D^0$ mesons is lower than the experimental data  at $p_T<4$ GeV/c, leading to an overestimated $\Lambda_c^+/D^0$ ratio in this range. Including the feed-down contributions from $D^*$ meson decays can significantly increase the $D^0$ yield, thereby reducing the $\Lambda_c^+ / D^0$ ratio at low $p_T$ and bringing the model into better agreement with the experimental data.

\begin{figure}[pht]
\includegraphics[width=0.45\textwidth]{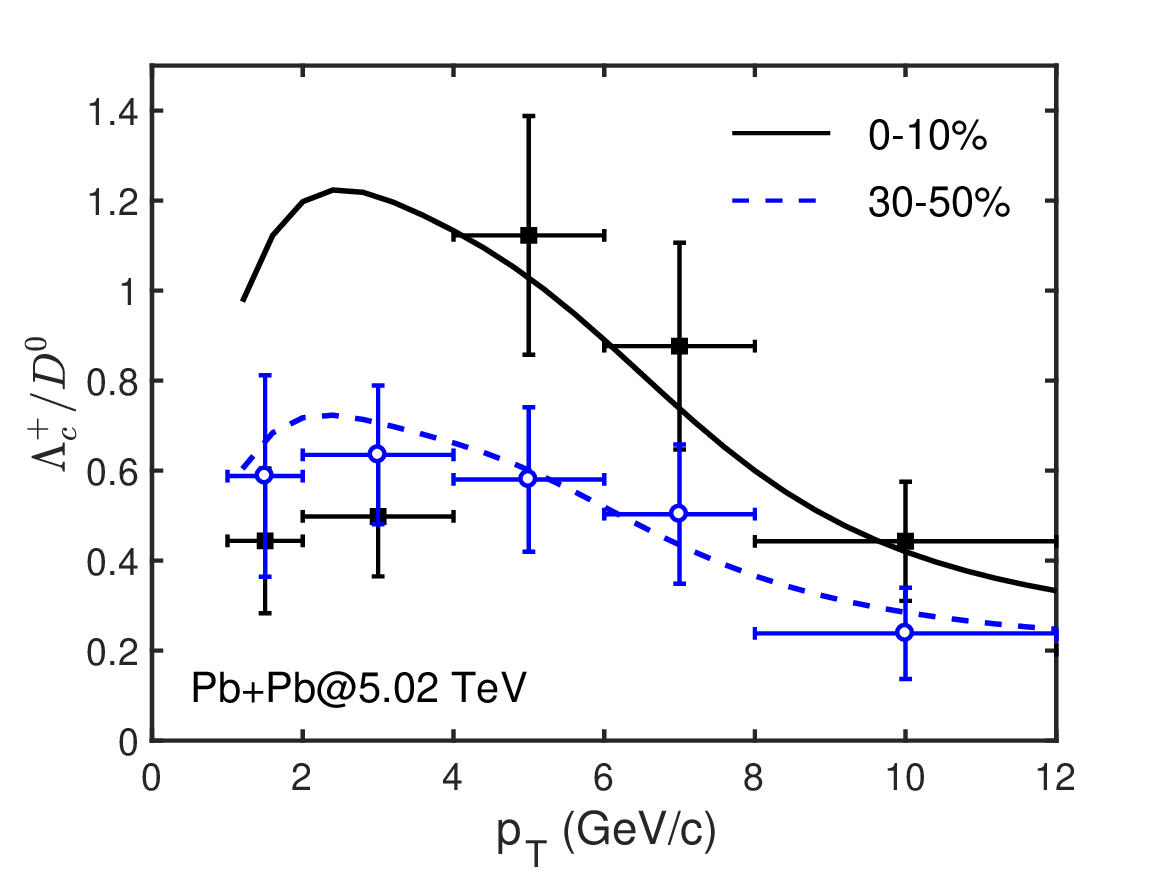}
\caption{(Color online) The $\Lambda_c^+/D^0$ yield ratio as a function of $p_{\rm T}$ in central and mid-central Pb+Pb collisions at $\sqrt{s_{NN}}=5.02$ TeV. The experimental data for  0–10\% (black solid squares) and  30–50\% (blue open circles) are taken from Ref. \cite{ALICE:2021bib}.}
\label{fig3}
\end{figure}

\begin{figure}[pht]
\includegraphics[width=0.45\textwidth]{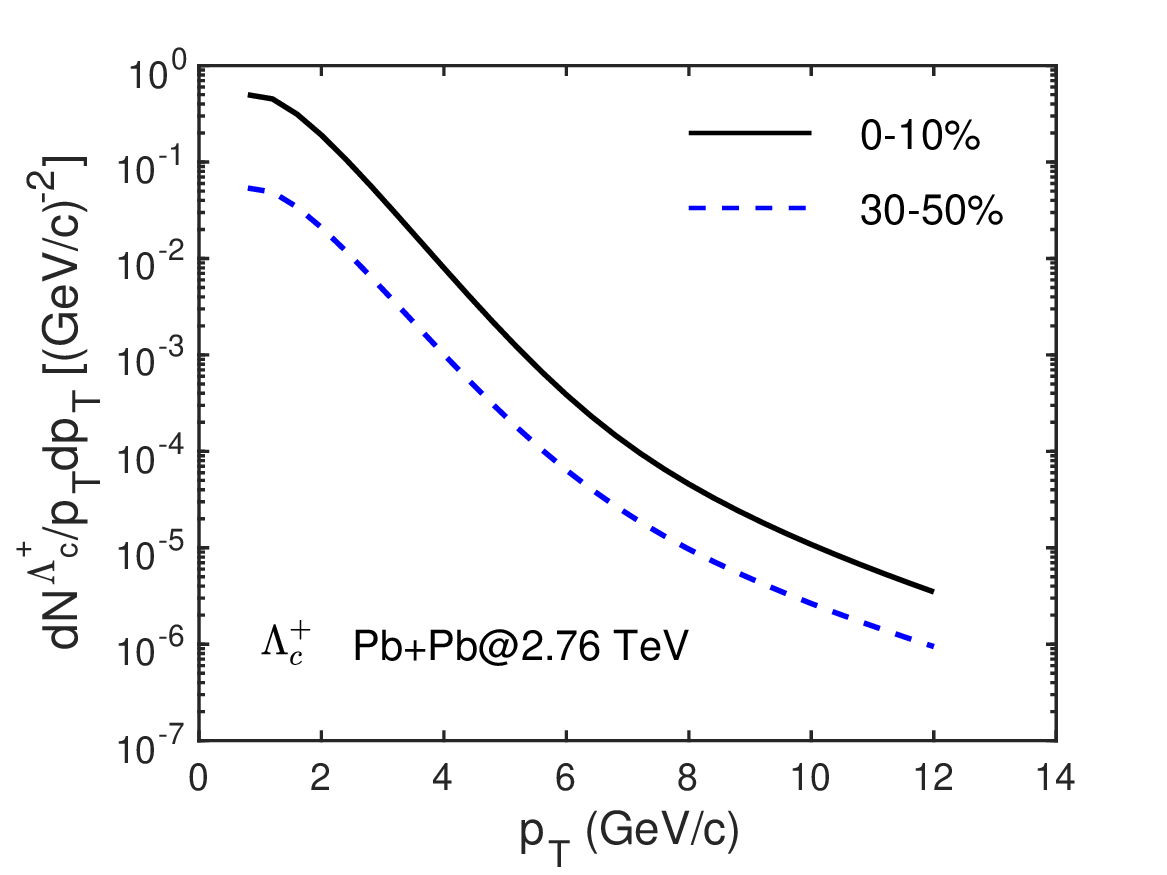}
\caption{(Color online) The predictions of transverse momentum spectra of $\Lambda_c^+$ baryons from the recombination model at centralities of 0-10\% and 30-50\% in Pb+Pb collisions at $\sqrt{s_{NN}}=2.76$ TeV. }
\label{fig4}
\end{figure}

\begin{figure}[pht]
\includegraphics[width=0.45\textwidth]{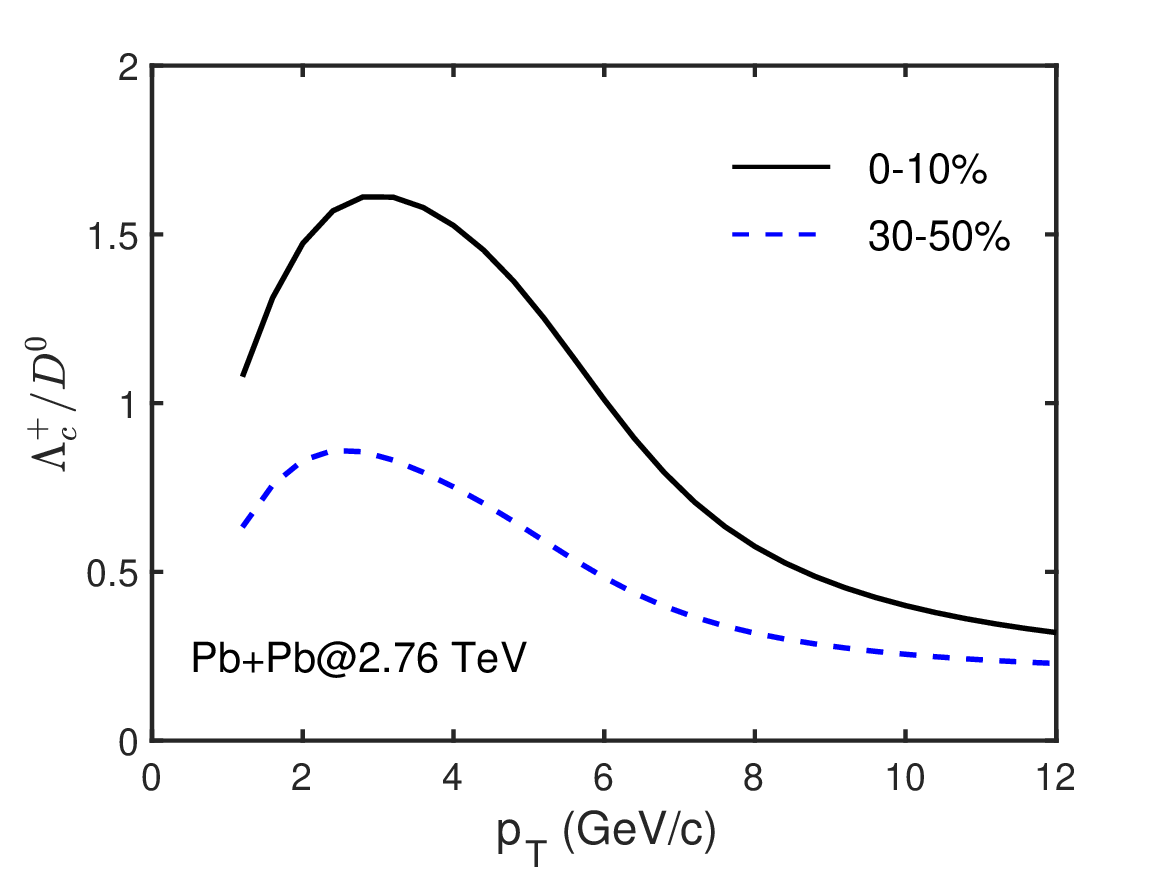}
\caption{(Color online) Predictions of the $\Lambda_c^+/D^0$ yield ratio from the recombination model at centralities of 0-10\% and 30-50\% in Pb+Pb collisions at $\sqrt{s_{NN}}=2.76$ TeV. }
\label{fig5}
\end{figure}

The statistical factor for $\Lambda_c^+$ baryons has been fixed by its transverse momentum spectra produced in Pb+Pb collisions at $\sqrt{s_{NN}}=5.02$ TeV as shown in Figs. \ref{fig1} and \ref{fig2}. All other parameters: $\gamma_0$ and $q_0$ in $\gamma$ factor, inverse slopes $T_q$ and $T_c$, and normalization factors $C_q$ and $C_c$ for Pb+Pb collisions at  $\sqrt{s_{NN}} = 2.76$ TeV, were determined in Ref. \cite{Zhang:2025ehd}. Therefore, we extend the investigation to $\sqrt{s_{NN}}=2.76$ TeV and present predictions for transverse momentum spectra of $\Lambda_c^+$ baryons and $\Lambda_c^+/D^0$ ratio in Figs. \ref{fig4} and \ref{fig5}, respectively, which can be tested once the experimental data become available.

\subsection{Second harmonic of azimuthal anisotropy of $\Lambda_c^+$}
In Refs. \cite{Zhang:2025ehd, Hwa:2012xy}, we investigated  the second azimuthal harmonic coefficient $v_2$ for $\pi$ and charmed mesons by including the effects of semihard partons, which can generate the minijets and give rise to $\phi$ anisotropy in the thermal component. It is worth extending the study to $\Lambda_c^+$ baryons. Since the detailed formulations have been presented in previous studies \cite{Zhang:2025ehd, Hwa:2012xy}, we focus on the key computational steps in this work. Let us use $\rho(p_T, \phi, b)$ to denote the $\Lambda_c^+$ distribution produced at midrapidity with impact parameter $b$, which is assumed to consist of three components: the base, ridge, and minijet contributions at low and intermediate $p_T$,
\begin{equation} 
\rho(p_T, \phi, b) = B(p_T,b) + R(p_T, \phi, b) + M(p_T, \phi, b).
\label{3.1}
\end{equation}
The component $B(p_T, b)$ is azimuthally isotropic, while $R(p_{T}, \phi, b)$ and $M(p_{T}, \phi, b)$ exhibit $\phi$ dependence. The base and ridge terms originate from recombination among thermal partons (TTT), while the minijet component arises from thermal-thermal–shower recombination (TTS). After averaging over $\phi$, we have
\begin{equation} 
\bar\rho(p_T, b) =B(p_T,  b) +\bar{R}(p_T, b) + \bar{M}(p_T, b).    
 \label{3.2}
\end{equation} 

The ridge component $R(p_T,\phi,b)$ incorporates azimuthal anisotropy that stems from the initial spatial eccentricity through the function $S_2(\phi,b)$, which converts spatial asymmetry into momentum anisotropy. The function $S_2(\phi,b)$ represents the segment of the medium surface through which a semihard parton can be emitted to produce a ridge particle at azimuthal angle $\phi$. When two nuclei of radius $R_A$ collide at impact parameter
$b$, the almond-shaped overlap has width and height given by 
 \begin{equation}
w=1-b/2, \hspace{0.5cm} h=(1-b^2/4)^{1/2},    
\label{3.3-1}
\end{equation}
where all lengths are normalized by $R_A$. The appropriate geometry that can describe the initial configuration is the ellipse
  \begin{equation}
(\frac{x}{w})^2+(\frac{y}{h})^2=1.  
\label{3.3-2}
\end{equation}
 Therefore, from the geometry of the ellipse, $S(\phi,b)$ can be determined by \cite{Hwa:2009rd},
\begin{eqnarray}	
S(\phi,b) &=&\int_{\theta_1}^{\theta_2} [w^2\sin^2\theta+h^2\cos^2\theta]^{1/2}d\theta  \nonumber \\
&=&h(b)[E(\theta_2,\alpha) - E(\theta_1,\alpha)],    
\label{3.3}
\end{eqnarray}	
where $E(\theta_i,\alpha)$ is the elliptic integral of the second kind with $\alpha=1-w^2/h^2$ and
\begin{eqnarray}
\theta_i = \tan^{-1} \left({h\over w}\tan\phi_i \right), \quad \phi_1 = \phi - \sigma, \quad \phi_2 = \phi + \sigma.  \nonumber \\    \label{3.4}
\end{eqnarray}
Therefore, the ridge component in Eq. (\ref{3.1}), which responds to the minijets through TTT recombination, can be expressed as
\begin{equation}
R(p_{T},\phi,b) = S(\phi,b)\bar{R}(p_{T},b).
\label{3.5}
\end{equation}
Similarly, the third component of $\rho(p_{T}, \phi, b)$ is written as
\begin{equation}
M(p_{T},\phi,b) = J(\phi,b)\bar{M}(p_{T},b),
\label{3.6}
\end{equation}
where $J(\phi,b)$ describes the azimuthal dependence of the minijet contribution and is defined as
\begin{equation}
J(\phi,b) = \widetilde{J}(\phi,b) \Big/ \left[\frac{1}{2\pi}\int_0^{2\pi} d\phi\widetilde{J}(\phi,b)\right].
\label{3.7}
\end{equation}
Since minijets are produced in any given event in unpredictable directions, the average $\phi$ distribution can have all terms in a harmonic analysis. Hence, the unnormalized function $\widetilde{J}(\phi,b)$ includes all harmonic components $\cos(n\phi)$, averaged over the angle $\psi_n$,
\label{3.8}
\begin{equation}
\widetilde{J}(\phi,b) = 1 + b\sum_{n=2}^{\infty} a_n\frac{n}{\pi}\int_{-\pi/2n}^{\pi/2n} d\psi_n \cos[n(\phi-\psi_n)],
\label{3.9}
\end{equation}
where $a_n$ are free parameters determined by fitting to the experimental data of $v_n$. Here we only consider the second harmonic of azimuthal anisotropy of $\Lambda_c^+$, so we can get \cite{Hwa:2009rd}
\begin{eqnarray}
v_2(p_T, b)={\left<\cos 2\phi\right>_S \bar R(p_T, b) + \left< \cos 2\phi\right>_J \bar M(p_T, b) \over \bar\rho^h(p_T, b)}  , \nonumber \\  
\label{3.10}
\end{eqnarray}
where
\begin{eqnarray}
\langle \cos2\phi \rangle_S &=& {1\over 2\pi} \int_0^{2\pi} d\phi \cos2\phi S(\phi, b),     \label{3.11} \\
\left< \cos 2\phi\right>_J&=&{1\over 2\pi} \int_0^{2\pi} d\phi \cos 2\phi J(\phi, b).  \label{3.12}
\end{eqnarray}

 \begin{figure}[pht]
\includegraphics[width=0.45\textwidth]{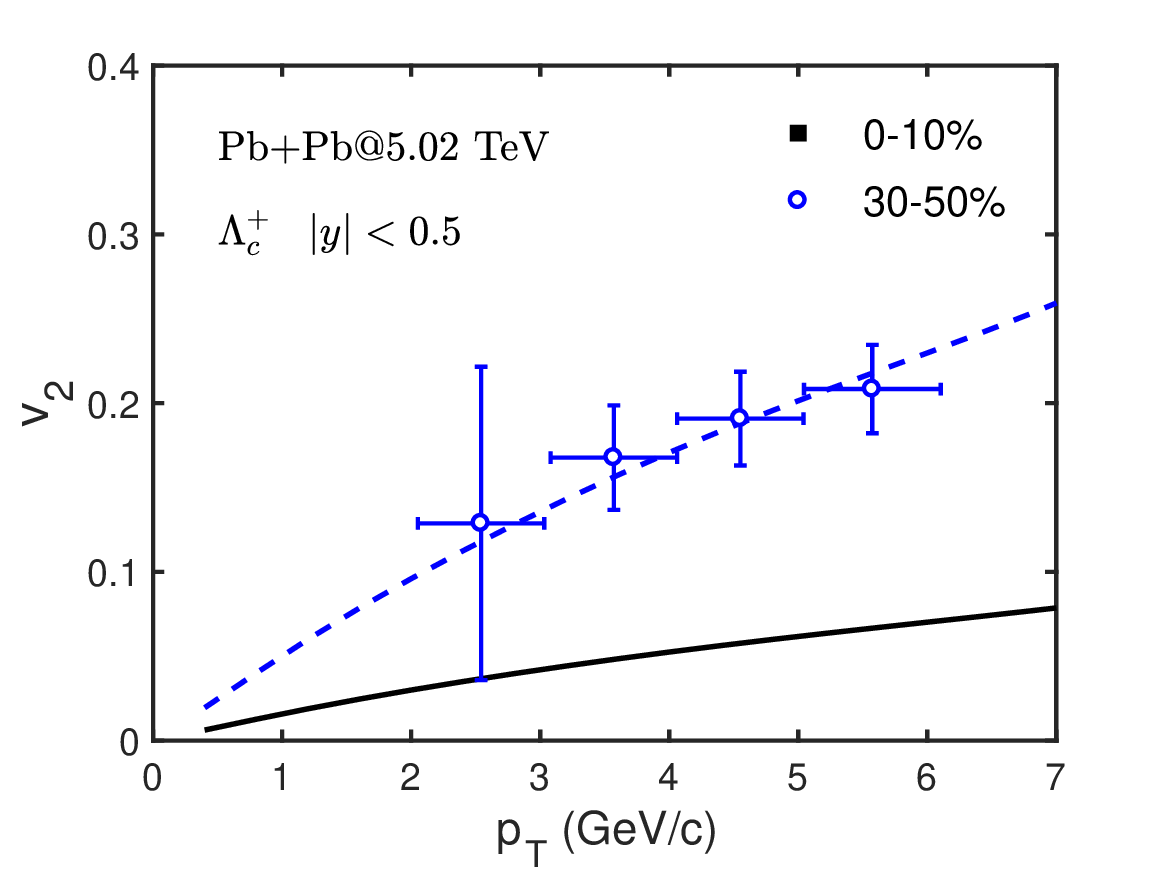}
 \caption{(Color online) The second harmonic coefficient $v_2$ of $\Lambda_c^+$ baryons at centralities of 0-10\% and 30-50\% in Pb+Pb collisions at $\sqrt{s_{NN}}=5.02$ TeV. The data are taken from Ref. \cite{DiCostanzo:2025zyb}.}
 \label{fig6}
\end{figure}

Figure~\ref{fig6} presents the second harmonic coefficient $v_2$ for $\Lambda_c^+$ baryons at centralities of 0-10\% and 30-50\% in Pb+Pb collisions at $\sqrt{s_{NN}}=5.02$ TeV. The model calculation reproduces very well the data for the centrality of 30-50\% from the ALICE Collaboration \cite{DiCostanzo:2025zyb}. The parameter $a_2$ is equal to 3.3. Unfortunately, the experimental data of $v_2$ for $\Lambda_c^+$ baryons at central collisions are not available yet. The theoretical prediction is shown in Fig. \ref{fig6}, which can be compared with experimental measurements in the near future. From the above study, we can conclude that the recombination model successfully describe the azimuthal anisotropy of mesons and baryons, which gives support to the essential dynamical role of minijets in the hadronization process \cite{Zhang:2025ehd, Hwa:2012xy}.

\section{summary}
\label{summary}
In this work, we investigate the production of $\Lambda_c^+$ baryons in Pb+Pb collisions at $\sqrt{s_{NN}}=5.02$ TeV with the quark recombination model that emphasizes the crucial role of minijets, which generate azimuthal anisotropy both through partonic energy loss in the hot medium and by producing shower partons that recombine with thermal partons. The model provides a unified description of three key experimental observables: the transverse momentum spectra of $\Lambda_c^+$  baryons, the enhanced $\Lambda_c^+/D^0$ yield ratio and the second harmonic coefficient $v_2$ for $\Lambda_c^+$  baryons. The successful reproduction of the centrality dependence of observables demonstrates that the interplay between charm quarks and the dense partonic medium is essential for understanding charmed baryon formation.
The results strongly support the dominance of quark recombination as the primary hadronization mechanism for charm quarks at intermediate transverse momenta. The inclusion of shower partons is indispensable for simultaneously accounting for the enhancement of the baryon-to-meson and the development of azimuthal anisotropy, thereby bridging the dynamics of soft and hard processes in heavy-ion collisions.
Furthermore, the model provides quantitative predictions for the transverse momentum distributions of $\Lambda_c^+$ baryons and the $\Lambda_c^+/D^0$ yield ratio in Pb+Pb collisions at $\sqrt{s_{NN}}=2.76$ TeV for future measurements.
This work provides compelling support for the picture that quark recombination is an effective hadronization mechanism for charm quarks in the hot and dense medium created at the LHC. This framework can be readily extended to make predictions for other charmed baryons, such as $\Xi_c$ and $\Omega_c$ baryons, which will provide further stringent tests of our understanding of heavy-flavor hadronization in the future.

\begin{acknowledgments}
H.Z. acknowledges financial support from the Key Laboratory of Quark and Lepton Physics at Central China Normal University (Grant No. QLPL2024P01).
\end{acknowledgments}

\end{document}